\documentclass[mathleft
]{an}
\usepackage{graphicx}
\usepackage{times}
\usepackage{amssymb}
\usepackage{amsmath}
\usepackage{booktabs}
\newcommand\kms{\ifmmode{\rm km\thinspace s^{-1}}\else km\thinspace s$^{-1}$\fi}
\usepackage{natbib}
\bibpunct{(}{)}{;}{a}{}{,}
\begin{document}

\Pagespan{789}{}
\Yearpublication{2006}%
\Yearsubmission{2005}%
\Month{11}%
\Volume{999}%
\Issue{88}%

\title{KIC011764567: An evolved {\it Kepler}-star showing substantial flare \\activity}

\author{M. Kitze\inst{1,3}\fnmsep\thanks{Corresponding author: 
\email{manfred.kitze@uni-jena.de}}, A. A. Akopian\inst{2}, V. Hambaryan\inst{1}, G. Torres\inst{4}, \and R. Neuh\"auser\inst{1}
}
\titlerunning{KIC011764567: Cyclic flare activity}
\authorrunning{M. Kitze et al.}
\institute{
Astrophysikalisches Institut und Universit\"ats-Sternwarte, Universit\"at Jena, Schillerg\"asschen 2-3, 07745 Jena, Germany
\and 
Byurakan Astrophysical Observatory, Byurakan, Armenia
\and 
Universit\"at Rostock, Institut f\"ur Physik, D-18051 Rostock, Germany
\and
Harvard-Smithsonian Center for Astrophysics, Cambridge, MA 02138, USA}

\received{30 May 2005}
\accepted{11 Nov 2005}
\publonline{later}


\abstract{%
We intensively studied the flare activity on the stellar object KIC011764567. The star was thought to be solar type, with a temperature of $T_{eff}\approx(5640\pm200)\:$K, $\log(g)=( 4.3\pm0.3)\:$dex and a rotational period of $P_{rot}\approx22\:$d \citep{brown}. High resolution spectra turn the target to an evolved object with $T_{eff}=(5300 \pm 150)\:$K, a metalicity of ${\rm [m/H]} = (-0.5\pm 0.2)$, a surface gravity of $\log(g) = (3.3 \pm 0.4)\:$dex, and a projected
rotational velocity of $v \sin i =( 22 \pm 1)\:$\kms. Within an observing time span of 4 years we detected 150 flares in {\it Kepler} data in an energy range of $10^{36} - 10^{37}\:$erg. From a dynamical Lomb-Scargle periodogram we have evidence for differential rotation as well as for stellar spot evolution and migration. Analysing the occurrence times of the flares we found hints for a periodic flare frequency cycle of $430 - 460\:$d, the significance increases with an increasing threshold of the flares equivalent duration. One explanation is a very short activity cycle of the star with that period. Another possibility, also proposed by others in similar cases, is that the larger flares may be triggered by external phenomena, such as magnetically interaction with an unseen companion. Our high resolution spectra show that KIC011764567 is not a short period binary star.}
  
\maketitle

\section{Introduction\\ }

Flares are sudden and abrupt brightenings of stars \citep{b1}. Though their relative amplitudes are much smaller than that of observed flares on late type M-dwarfs, the total amount of energy released in flares on solar type stars can exceed $10^{33}-10^{35}\:$erg. A first list of solar like flare stars was presented by \cite{b2}. Four dozen flares have been reported in \citet{b2,b3,b4}. According to \cite{b1}, flares can occur on single, middle-aged and slowly rotating main-sequence stars of spectral type F8 to G8 (``solar type stars''). The similarity of stellar flares to solar flares suggests that stellar flares can arise from magnetic effects. For the explanation of flare phenomena the model of ``hot Jupiter'' was involved (\cite{b1} and references therein). In this model the author suggests the existence of a strong magnetic field that connects the star to an orbiting planet. 

The first flare detection lists are collections of heterogeneous data that does not allow for proper statistical analysis and conclusions. The situation has radically changed after {\it Kepler} observations. Designed to detect transiting exoplanets, the observatory {\it Kepler} is also the ideal tool for the study of variable stars and the diverse  phenomena of variability. Further the continuously sampled {\it Kepler} observations improve the chance to simultaneously detect flares, and the transiting giant planet, which might be responsible for those flares. This gives the principal possibility to directly check the hypothesis of the ``hot Jupiter'' model.

Considering the data obtained by the observatory {\it Kepler}, \cite{b5} have registered 365 flares on 148 stars, including some from slowly rotating solar type stars, from about 83,000 stars observed over 120 days. The non detection of transiting planets in solar type flare stars led to the assumption, that ``hot Jupiters'' around solar type flare stars are rare. Later, \citet{shiba} confirmed this result using the observational data for $500\:$d. 

A more convincing argument against the ``hot Jupiter'' model would be to deny a connection between a hypothetical close companion and the flare activity of the host star from investigating the flares occurrence times. In case of a close companion, either in an eccentric or a circular orbit, one might expect magnetically interaction, hence a variable flare activity, either due to a preferred orbital phase or a preferred line-of-sight.

From exploring a small sub-sample of stars that most frequently flared within the list of \cite{b5}, \cite{b6} has found a statistically significant change of the flare frequency for the star KIC007264976. However, the data presented in \citet{b5} did not allow to reveal cyclic variability, due to the relatively small number of flares. Later, using the data of \citet{shiba}, \citet{akopian2} has found statistically significant changes in the frequency of flares for five stars. Mostly the moments of change of frequency are accompanied by sudden changes in the behaviour of the stellar brightness. Then the brightness of a star becomes irregular for a short time, with a significant decrease in the amplitude of brightness. Availability of possible periodicity of the flares frequency with a period equal to the period of axial rotation of the star or the orbital motion of a companion was also examined. It was shown, that the flare frequency of the star KIC007264976 has a period of $134\:$d, which coincides with the possible period of orbital motion \citep{wich,akopian2}. A period of $5.2\:$d equal to the rotational period was revealed in case of the star KIC010422252 \citep{akopian2}. 

In this study we present results concerning to the star KIC011764567. We investigated spectroscopically, if the star is really solar like (Sec. 2.1), since the star has a high potential to counterfeit the flare frequency statistics and the flare energy distribution of intrinsic activity of isolated solar type stars. In Sec. 2.2 we performed a photometric analysis, including the investigation of rotational features, the detection and characterisation of flares, using our spectroscopic results, and the investigation of any correlation between flares and the rotation. Since the large amount of detected flares gives the unique chance to study the flares occurrence times, we investigated, if the flares can be interpreted with a Poissonian-distribution (equally distributed), have a preferred occurrence phase (irregular occurrence) or show a cyclic behaviour (periodic occurrence). Therefore we applied two independent methods to the data (Sec. 2.3).

For analysis, we involve more complete {\it Kepler} data available to date, which cover the whole period of observations.

\section[Data analysis \& Results]{Data analysis \& Results}
\subsection[Spectral analysis]{Spectral analysis}
KIC011764567 is a presumable solar like slowly rotating main sequence star \citep{b5}. Data of different authors, given in Tab. \ref{tab1}, bare a wide spread of the parameters $T_{\text{eff}}$ and $log\text{g}$. \citet{wich} have taken CAFE spectra of the star from Calar Alto. The non-detection of Li I $\lambda 6707$ indicates the star to be probably old. Further the measured $v \sin i$ of $(21\pm3)\:$\kms \ is too large for a main sequence star of the given rotational periods. \citet{wich} suggested that the star may be either evolved or be a binary. 
\begin{table}
\begin{center}
\caption{Data of KIC011764567 from different studies: [1] \citep{brown,b5}, [2] \citep{pinsonne}, [3] \citep{wich}.}
\begin{tabular}{c|l|c|c}
\hline reference & $T_{eff} \ [K]$ & $Period \ [d]$ & $log(g)$ \\\hline
\hline
[1] & $5238 \pm200$ & $22.7$ & $4.38\pm0.40$ \\\hline
\hline
[2] & $5480 \pm66$ & --- & --- \\\
    & $5586 \pm130$ & --- & --- \\\hline
\hline
[3] & $6100 \pm300$ & $20.5$ & $4.5\pm 0.5$ \\\
 & $5640 \pm240$ & $20.5$ & $3.5\pm 0.5$\\
\end{tabular}
\label{tab1}
\end{center}
\end{table}

In this work KIC011764567 was observed spectroscopically at the Harvard-Smithsonian
Center for Astrophysics with the Tillinghast Reflector Echelle
Spectrograph \citep[TRES;][]{Furesz:08} attached to the 1.5\,m
telescope at the F.\ L.\ Whipple Observatory on Mount Hopkins, Arizona
(USA). 11 echelle spectra covering the wavelength range
3860--9100\,\AA\ in 51 orders were gathered between 2015 March and May
at a resolving power of $R \approx 44\:000$. They were reduced and
extracted using standard procedures as described by
\cite{Buchhave:10}.  The signal-to-noise ratios average about 15 per
resolution element of 6.8\,\kms, at a mean wavelength of
5200\,\AA. Radial velocities were computed by cross-correlation
against a synthetic template selected to match the stellar
properties. The template parameters were optimized by running
extensive grids of cross-correlations against a large library of
calculated spectra based on model atmospheres by R.\ L.\ Kurucz
\citep[see][]{Nordstrom:94, Latham:02}. The best match was found for a
temperature of $(5300 \pm 150)$\,K, a metallicity of ${\rm [m/H]} = (-0.5
\pm 0.2)$, a surface gravity of $\log(g) = (3.3 \pm 0.4)\:$d, and a projected
rotational velocity of $v \sin i =( 22 \pm 1)$\,\kms. The resulting
heliocentric radial velocities are reported in Tab. \ref{tab_rv} along with
their uncertainties, and are on the system defined by IAU standards.
The measurements indicate no significant variation over the period of
observation. The weighted average radial velocity is $(-2.26 \pm
0.23)$\,\kms. This appears to rule out the presence of a stellar companion of significant mass in orbits with periods under about 70 days.
\begin{table}
\begin{center}
\caption{Heliocentric radial velocities of KIC011764567}
\begin{tabular}{c|c|c}
\hline
 HJD-2.400.000 &   RV  &    err\ \\
\hline
57113.9996 &  -2.58 &   0.72\ \\
57117.9682 &  -1.59 &   0.47\ \\
57119.9037 &  -2.56 &   0.93\ \\
57121.9672 &  -1.54 &   0.87\ \\
57139.9061 &  -2.66 &   0.61\ \\
57143.9689 &  -2.75 &   0.88\ \\
57146.9224 &   0.07 &   1.61\ \\
57152.9680 &  -3.73 &   1.17\ \\
57168.9497 &  -2.56 &   0.70\ \\
57174.8436 &  -2.50 &   0.89\ \\
57184.8794 &  -2.55 &   1.02\ \\
\hline
\end{tabular}
\label{tab_rv}
\end{center}
\end{table}

The surface gravity corresponds to an evolved star. A typical absolute
visual magnitude for a giant of this temperature implies a distance of
the order of 3 kpc. Estimates of the reddening at this distance based
on extinction maps yield a value of $E(B-V) \approx 0.10$.  Colour
indices for KIC011764567 based on standard photometry were gathered
from the literature, and corrected for reddening. Use of five indices
and the color/temperature calibration of \cite{Ramirez:05} results in
an average temperature of $(5370 \pm 130)\:$K, in agreement with the
spectroscopic value.

\subsection{Photometric analysis}

\begin{figure}
 \includegraphics[width=5.8cm,angle=-90]{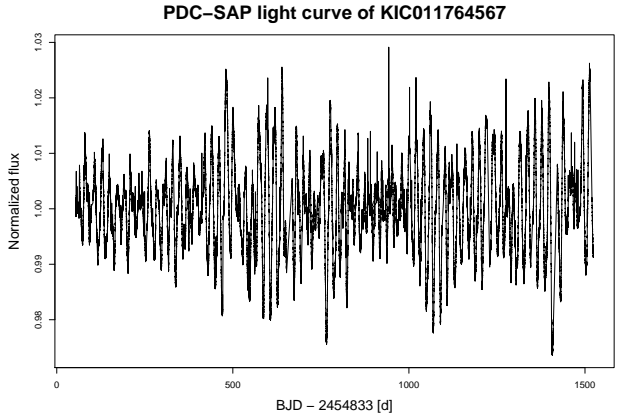}
 \caption{PDC-SAP light curve of KIC011764567 for quarter 0 to quarter 17 without the flares. The light curve pattern shows a well defined periodic behaviour. This stellar variation is obviously dominated by stellar spot modulation.}
 \label{pdcsap}
\end{figure}

Analysing data of quarters $0,1$ and $2$ of {\it Kepler} observations, \cite{b5} detected 11 flares on the star KIC011764567, that were interpreted as superflares in the solar type context. We analysed data of the quarters $0-17$ which were retrieved from the Mikulski Archive for Space Telescopes (MAST)\footnote{http://archive.stsci.edu/kepler/} and found additional 139 flares. A wavelet based filter was applied to first detrend the PDC-SAP flux data for the present stellar variability (Fig. \ref{pdcsap}). When comparing the original data with the wavelet representation of the signal in only lower frequency pass bands, flares induce a strong signal in the residuals as the wavelet fit cannot reproduce the sharp and high frequent rising part. To test our method we randomly placed synthetic flare events of different shape and amplitude into 3000 simulated and real quarterly {\it Kepler} data segments and searched for them afterwards. We were able to detect $> 80\:\%$ of such short time events with amplitudes $> 0.05\:\%$ and $> 90\:\%$ with amplitudes $> 0.08\:\%$ for the brightness of KIC011764567 ($K_{p}=13.2\:$mag, Brown et al. 2011), which is in the order of the photometric errors. To distinguish between flares and other types of variability we fitted a mathematical expression \citep{kitze} - a simplification of the complex behaviour of a flare event to describe the sharp rising and the smoothed e-folding relaxation part in the optical wavelength range - to that time points, where the residuals exceed $3\:\sigma$ and further characterized the so identified ``real'' flares. In some particular cases the detected flares have a more complicated shape, so that we also introduced a double peak model. For the double peak flares, we used the highest amplitude peak as reference for the flares peak time. The end of a flare event was determined with F-Test statistics with a commonly used 95\% confidence interval.

\begin{figure}
 \includegraphics[width=5.8cm,angle=-90]{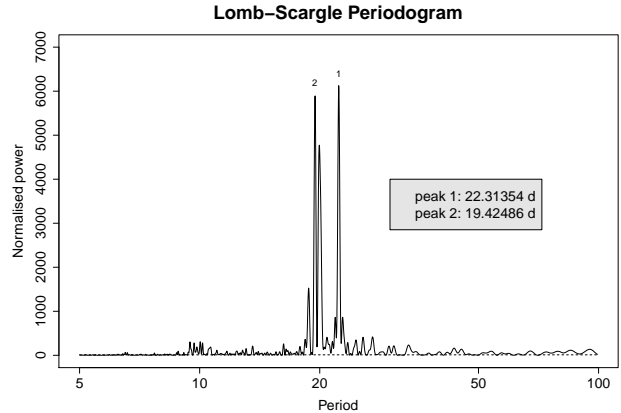}
 \caption{Lomb-Scargle Periodogram of the combined light curve of quarter 0 to quarter 17 for KIC011764567. There are 2 well separated major peaks, ranging from $19$ to $22\:$d.}
 \label{lomb}
\end{figure}

\begin{figure*}
\centering
 \includegraphics[width=.58\textwidth ,angle=-90]{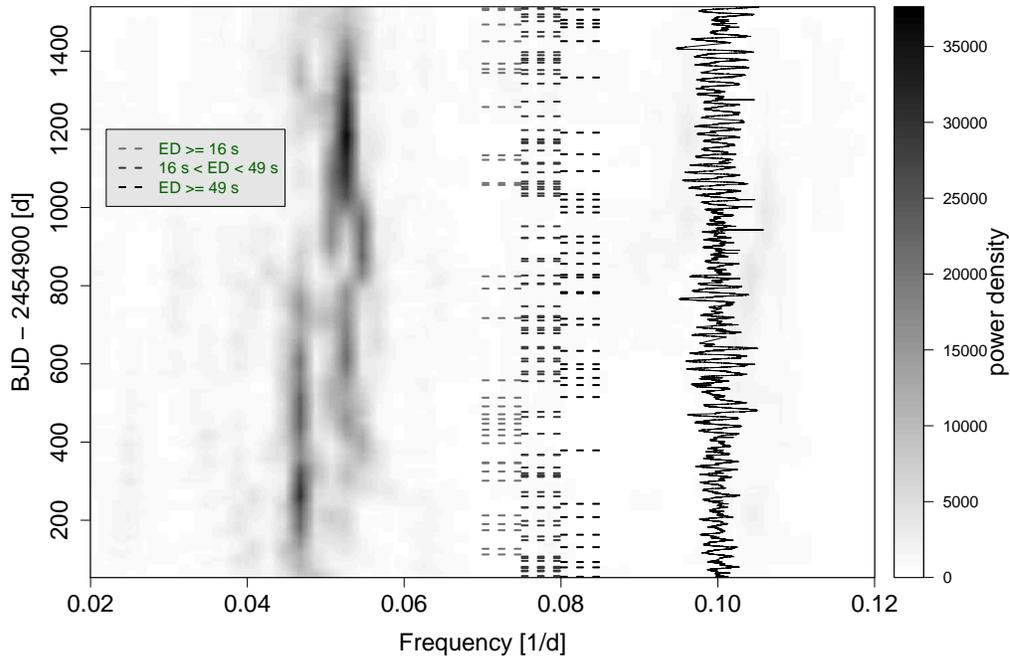}
 \caption{Dynamical Lomb-Scargle plot of the frequency vs. time. For a comparison the full light curve was projected to the diagram (solid line). The rotation frequency distribution is presented on a grey shaded scale (left), referred to the significance of the LS peak. Regarding to the power densities the observed effective frequency changes during the observing time span. This is dependent on the position and evolution of stellar spots. Additionally the occurrence times of flares of different equivalent durations are illustrated as horizontal lines in the middle. The smallest and largest flares show some larger time gaps, that are delayed in time.}
\label{scargle}
\end{figure*}

The PDC-SAP light curve of KIC011764567 contains a typical sign of stellar spot modulation (Fig. \ref{pdcsap}). In the Lomb-Scargle (LS) periodogram of the combined light curve of quarter 0 to quarter 17 there are well separated major peaks at around $22.31\:$d and a triple peak centred at $19.42\:$d (see Fig. \ref{lomb}). In a dynamical Lomb-Scargle representation with a time resolution of $20\:$d ($\approx 1\:P_{rot}$) and an analysing window of $200\:$d ($\approx 10\:P_{rot}$, see Fig. \ref{scargle}), we see that - dependent on the effective spot latitude - the detectable rotational period significantly changes during the $4\:$yr of observations. This is a hint for differential rotation. The prominent periods of $22.31\:$d and $19.42\:$d are relocated in Fig. \ref{scargle} at around $0.045\:\text{d}^{-1}$ and $0.051\:\text{d}^{-1}$.
 
Fig. \ref{scargle} additionally presents the location of the flares for different equivalent durations in the time domain. Regarding to these occurrence times, flares in a certain subsample can disappear over a large time span up to $200\:$d, compared to the averaged waiting time of $\approx 10\:$d (simply the observation time divided by the number of detected flares). It is interesting to see, that the large time gap of the smallest flares at around $800 - 1000\:$d is strongly connected to a low amplitude phase of the time signal. We further see, that the time gaps of the smallest and largest flares are systematically delayed in time. However not least due to the presence of stellar spots we expect the flares to be released energy phenomena from active regions.
\begin{figure*}
  \begin{center}
    \vspace{-0.2cm}
    \vbox{
      \hbox{
        \includegraphics[width=6.0cm,angle=-90]{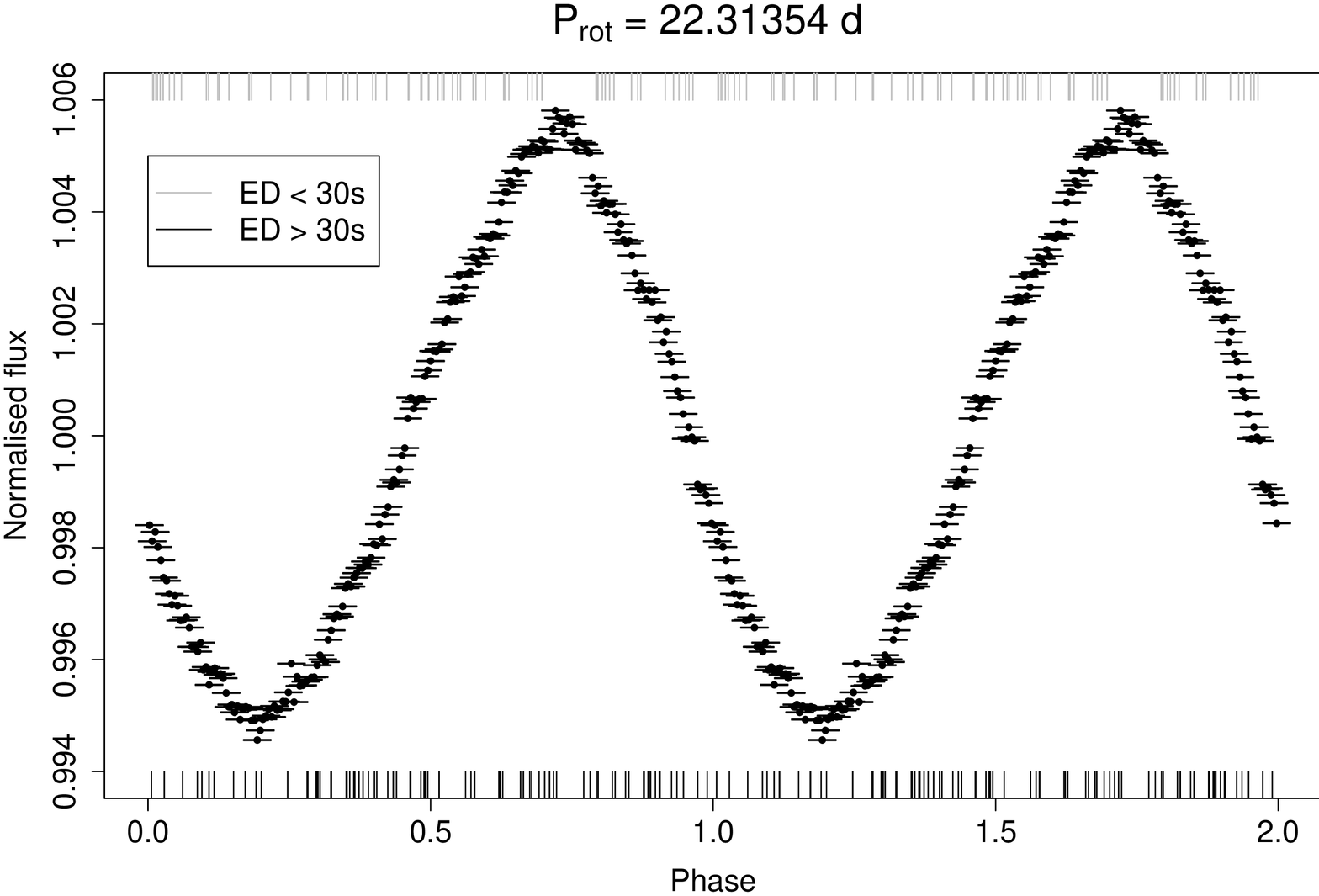}
        \includegraphics[width=6.0cm,angle=-90]{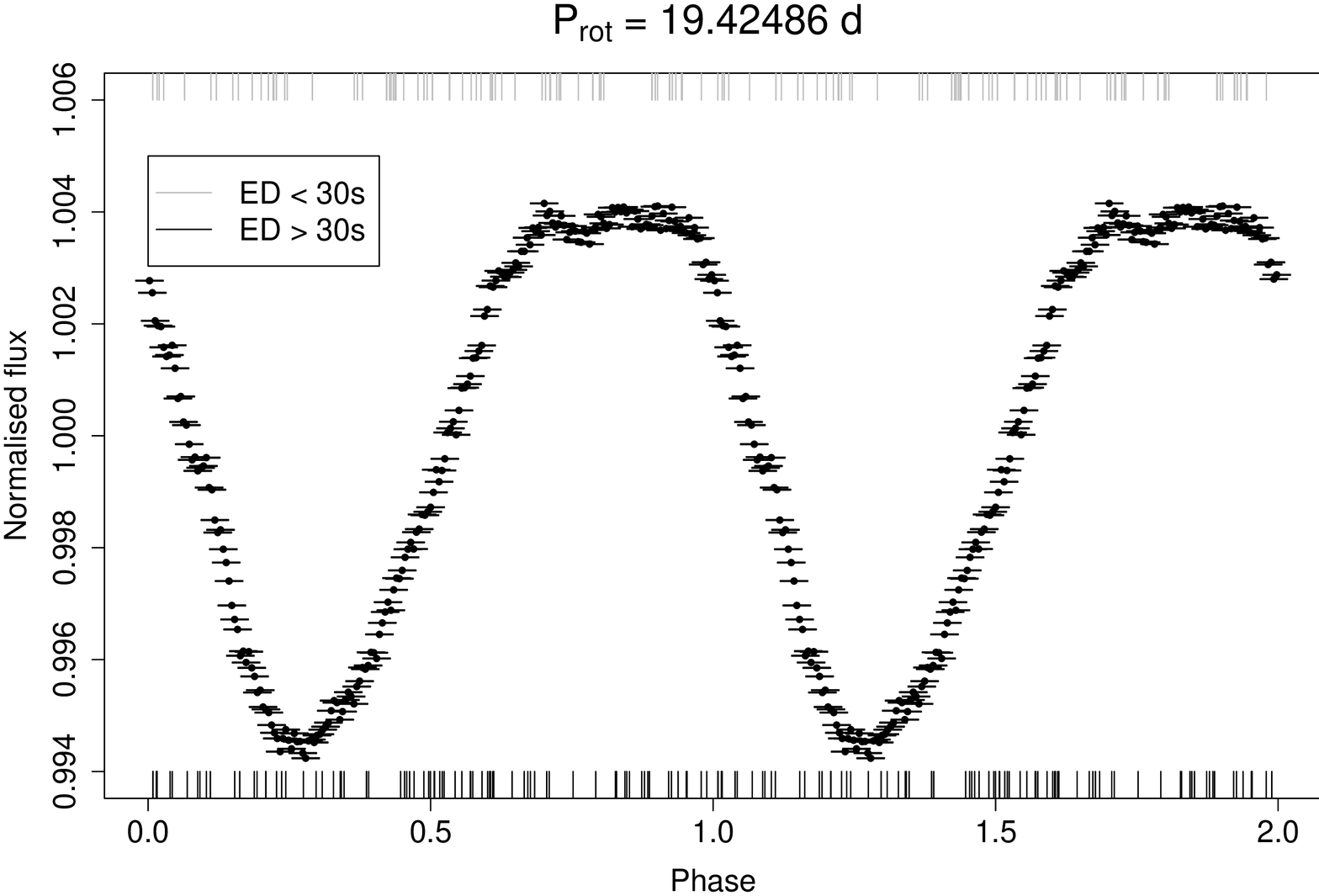}
      }
    }
    \vspace{-0.2cm}
    \caption{Phasefolded light curves for the most dominant peak frequencies regarding to Fig. \ref{lomb}. Additionally the folded occurrence times of the detected flares are presented for equivalent durations $<> 30\:$s. There is no strong correlation between the flares and the rotational period.}
    \label{diffrot}
  \end{center}
\end{figure*}

We have folded the light curve and additionally the time series of the flares with lower and higher equivalent durations (Fig. \ref{diffrot}) into the most dominant periods of the LS periodogram to check if the flares have a preferred occurrence phase during the rotation. We used a threshold of $30\:$s equivalent duration to fragment the flares in two equal sized groups. There is no preferred occurrence phase for flares of long and short equivalent durations.

Supposing an equatorial rotation period of $P_{equ}\approx (19.4\pm1.5)\:$d as indicated by the Lomb-Scargle analysis and a projected surface velocity of $v \sin i=(22\pm1)\:$\kms  a lower limit for the radius can be estimated to $R_{min}=(8.45\pm1.04)\:R_\odot$. $R_{min}$ and $T_{eff}$ from spectral analysis were used to estimate the minimum bolometric flare luminosities and energies. From uncertainty of $R_{min}$ and $T_{eff}$ we reach an uncertainty of $36\%$ for $L_{bol}^{min}$ and $E_{bol}^{min}$ which is more than two times smaller than from data of the Kepler Input Catalogue \citep{brown}. Owing to the similarity to solar flares, we assumed that the optical part of the flare energy distribution can be described by black body radiation of $10000\:$K \citep{kretzschmar}. We calculated bolometric flare energies and luminosities from distance modulus - using the values from Sec. 2.1 and the photometric data from 2MASS \citep{cutri,skrutskie} (bolometric correction (BC) was done in K-Band, using BC coefficients from \citet{bessell}) - and from the {\it Kepler} filter response \citep[identical method]{shiba}. Both results are in agreement within the error bars.

\begin{figure}
\includegraphics[width=5.8cm, angle=270]{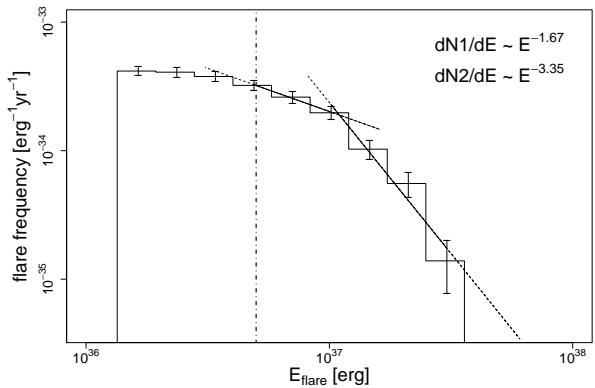} 
\caption{Flare distribution in terms of frequency per erg per year. The dash dotted vertical line is connected to the $95\%$ detection limit. The high energy part was fitted with a power law having one break at $10^{37}\:erg$. The estimated power law indices of $ -3.35\pm0.35$ and $-1.67\pm0.06$ bracket the index of solar flares \citep[$\alpha \approx -2.0$]{schrijver}}
\label{flares}
\end{figure}

Fig. \ref{flares} illustrates the distribution of the flare frequency over the flare energy for all detected flares in terms per erg per year. The energy bin size is chosen to be equidistant on a logarithmic scale and the error bars correspond to the square root of each energy bin sample. The detection limit at around $5*10^{36}\:$erg (vertical dash dotted line in Fig. \ref{flares}) corresponds to a detection rate of $95\%$. The higher energy part was fitted with a power law having one break at around $10^{37}\:$erg. The power law indices of $( -3.35\pm0.35)$ and $(-1.67\pm0.06)$ bracket the behaviour of solar flares \citep[$\alpha \approx -2.0$]{schrijver}.
  
Finally we checked the astrometric signal during the flares to check if any background star, that is well separated from KIC011764567 but still inside the {\it Kepler} pixel mask of the target, might be responsible for the flares. Therefore we calculated the photometric barycenter for each time stamp using a simple centroid algorithm and corrected the signal for the unstable Kepler pointing \citep{kitze}. The flares do not change the photometric barycenter significantly, so that there is no further link to a background star.

\subsection[Investigation of the flares occurrence times]{Investigation of the flares occurrence times}
Within the framework of existing concepts, one can expect a periodic variability of the flare frequency either due to an intrinsic short term activity cycle or magnetically interaction with a planetary or stellar companion around the primary star \citep{b1}. Thus, two methods were used, in one of which a periodic function with unknown parameters was used.

To identify a periodic signal and to specify the frequency precisely, we performed an analysis 
\citep[e.g. Bayesian analysis of the search, estimate and testing hypothesis][hereafter referred to as the GL method]{GL1992} of the occurrence times of the flares. The GL method for timing analysis first tests if a constant, variable or periodic signal is present in a data set, consisting of the occurrence times of events, when we have no specific prior information about the shape of the signal.

In the GL method, periodic models are represented by a signal folded into a trial frequency with a light curve shape as a stepwise function with $m$ phase bins per period plus a noise contribution. With such a model we are able to approximate a phase-folded light curve of any shape. Hypotheses for detecting periodic signals represent a class of stepwise, periodic models with the following parameters: trial period, phase, noise parameter, and number of bins ($m$). The most probable model parameters are then estimated by marginalization of the posterior probability over the prior specified range of each parameter. In Bayesian statistics, posterior probability contains a term that penalizes complex models (unless there is no significant evidence to support that hypothesis), so we calculate the posterior probability by marginalizing over a range of models, corresponding to a prior range of number of phase bins, $m$, from 2 to 12.

After having strong support to the hypothesis for the presence of a periodic signal, the frequency and its uncertainty is estimated by using the posterior probability density function \citep{GL1992}.

Independently to the GL-method, the analysis, described in \citet{b6,akopian2} was performed to the flares occurrence times, where a cyclic intensity function is fitted to the data \citep{b7, b10}. 

 \begin{figure}
\includegraphics[width=.48\textwidth]{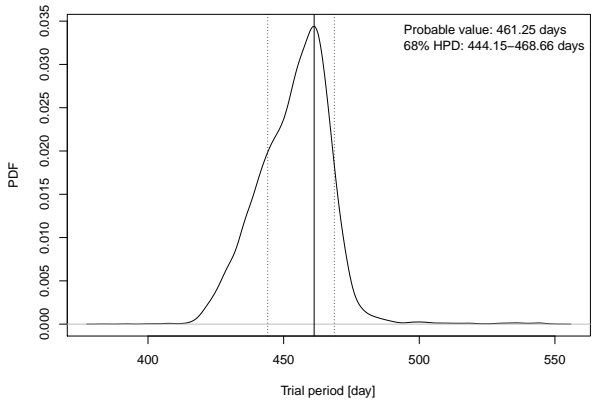}
\caption{Bayesian posterior probability density function of the 
trial periods by application of the GL method for periodicity detection for only a subsample of flares with $ED > 30\:$s. The significance of the periodic signal has increased, compared to the period detection for all flares.}
\label{difdistr}
\end{figure}

Using the methods, we found hints for a variable flare frequency. We applied the Gregory-Loredo (GL) Bayesian method ({\it method 2}) for periodicity detection in the frequency range of 0.0015$-$0.035~day$^{-1}$ to the flare registered times of KIC011764567 and detected a period of $(432.11^{+29.01}_{-14.48})\:$d. It is very interesting to note that this value coincides with high precision with the value of minimal common period ($433.26=19.42\cdot22.31$) of the two prominent periods of differential rotation, mentioned above ($19.42$d and $22.31$d ). It is difficult to say, if there is any physical basis under this coincidence. However it draws a connection of the value of $432.11\:$d and the occurrence times of the flares to the rotation of the star.

The frequency range was chosen to have at least 2 full periodic cycles within the observing time span on the one hand and not to break Nyquist`s rule on the other hand. Note, that the significance of the period detection increases when only analysing flares with larger equivalent durations. For flares with $ED > 30\:$s we get a period of $(461.25^{+07.41}_{-17.10})\:$d (Fig. \ref{difdistr}) where the uncertainty range is reduced by a factor of 2. Both methods, applied for period detection give similar results. The significance of the periodic signal is still too low for definite confirmation. However for statistical reasons a periodic model describes the observations better than a constant rate. So is the natural log of maximum likelihood ratio of these hypotheses (periodical frequency / constant frequency) equal to 4.9.

\section[Discussion]{Discussion\\}
Recently \citet{aulanier} have discussed, that our present sun is probably not able to generate flares larger than $O(10^{33})\:$erg. In this context it is controversial discussed, if the increase in the concentration of $^{14}C$ in tree rings from the years AD 774/775 were evoked by a large solar flare of $10^{35}\:$erg \citep{miyake,usoskin}, though there is no historical record rendered during this time \citep{neuhaeuser,rne}. 

Flares of $10^{33}\:$erg and orders of magnitudes larger have been observed several times on presumable solar type stars. Currently known frequency statistics, based on these detections \citep{b5,shiba} predict flares of $10^{35}\:$erg on average every $5000\:$yr for a single sun-like star. \citet{kitze} could show, that not all the detected flares in this energy range stem from the solar type stars. They showed, that KIC007133671 is probably not the source for the one detected flare which is instead generated by a different source within the Kepler Pixel mask of the target. KIC011764567 was a further, quite promising sun like flare star with presumed similar rotation, similar temperature and probably similar age as the sun. The large $v \sin i $ of $(22\pm1)\:$\kms and the low surface gravity of $(3.3\pm0.4)$ from our spectral analysis turn the object to an evolved star. Hence these flares cannot be used in further frequency statistics of intrinsically stellar activity of solar type stars. 

``Hot Jupiters'' around solar like stars have periods in the order of several days. Assuming an isotropic radiation for evolved objects of $R_{min}=(8.45\pm1.04)\:R_\odot$ and similar spectral type, the orbital period of ``Hot Jupiters'' can increase to tens of days up to around hundred days following the Sudarsky classification \citep{sudarsky}. Orbital periods of $< 70\:$d could be excluded from our spectral analysis, but only for stellar companions, not for sub-stellar companions. With the GL method we can cover a period range of $\approx 30 - 670\:$d where we detected a most probable flare frequency cycle of $(432.11^{+29.01}_{-14.48})\:$d for all flares. This is significantly longer ($> 4$ times) than expected for ``Hot Jupiters'' around this spectral type. The value of $432.11$d coincides with the value of minimal common period $433.26$ of the two prominent periods of differential rotation. This is a strong argument in favour of a differential rotating surface and a connection of the flares to the magnetically active surface of the star.

We want to emphasize, that it is an important task to study numerous more such solar like flare stars in detail, since new results would have a high impact to the low number statistics. These statistics are still based on only 6 investigated {\it Kepler} quarters and they do not take into account the age of the stars. Further \citet{wich} have shown recently, that there is a wide spread in temperatures between the KIC \citep{brown} and from high resolution spectra, leading to an uncertainty in spectral types of several subclasses. However, KIC011764567 is still an interesting target to study the periodic behaviour of the flares occurrence frequency. 
\section[Summary]{Summary\\}
In this work KIC011764567 was observed spectroscopically to better restrict its spectral type. As a result of this analysis KIC011764567 has turned to be a yellow giant. Consequently its flares cannot be used in frequency statistics of intrinsically stellar activity of single solar type stars any more. Nevertheless the flare activity of KIC011764567 and its connection to the stellar rotation, spot activity and a hypothetical companion was intensively studied, since yellow giants are rare objects. We investigated the rotational period as a function of time. From the dynamical Lomb-Scargle diagram we have strong support for the existence of differential rotation and resulting spot migration as the rotation frequency changes significantly. This is an important link to the far more numerous active yellow dwarfs and could give useful suggestions for theoretical studies of Dynamo models. 

We detected 150 flaring events in our data. Comparing the flares occurrence times with the time signal of stellar spot modulation, we have evidence, that the flares are connected to the spot activity. 

We used the occurrence times of the 150 detected flares for further analysis to check, whether the time events have a proper Poissonian distribution, vary their frequency with time or have a periodic background. From two independent methods, using a set of cyclic intensity functions \citep{b7, b10} and the GL method \citep{GL1992}, we have support, that the flares frequency changes with time. Especially the stronger flares with $ED > 30s$ increase the probability for the presence of a change in frequency. The most probable period was estimated to be in the range of $430-460\:$d, using maximum likelihood estimates and posterior probabilities density function. However the odds ratio of the GL method is still to low to definitely confirm the periodic signal. For reasons of a much longer period than expected for a ``hot Jupiter'' and the related weaker field strength of magnetic interaction, the detected period of $430-460\:$d more likely displays an unknown intrinsic activity cycle of the star than a connection to an unseen companion. This is supported by the coincidence of the period with the minimal common period of the dominant periods of differential rotation.

The significance of the periodic signal and the parameter space can be increased by using more time events and a longer observing time span, hence more observations have to be necessarily done for this target. Currently, using 79 flares with $ED > 30\:$s we are not able to search for periods $< 36\:$d, referred to Nyquist`s rule.

\section*{Acknowledgments}
MK and RN would like to thank DFG in SPP 1385 project NE515/34-1,34-2 for financial support. VVH would like to thank DFG in SFB TR7 for financial support. MK would also thank Ronald Redmer and DFG in project RE 882/12-2 for financial support.


\begin{thebibliography}{}
\bibitem[\protect\citeauthoryear{Akopian}{2013}]{b6} Akopian A.A., 2013, Astrophysics, 56, 488, translation from Astrofizika (in russian), 56, 537

\bibitem[\protect\citeauthoryear{Akopian}{2015}]{akopian2} Akopian A.A., 2015, Astrophysics, 58, 62, translation from Astrofizika (in russian), 57, 75 

\bibitem[\protect\citeauthoryear{Ambartsumian}{1969}]{b8} Ambartsumian V.A., Stars, Nebulae, Galaxies, 1969, Yerevan

\bibitem[\protect\citeauthoryear{Aschwanden}{2011}]{b9} Aschwanden M., Self-Organized Criticality in Astrophysics, 2011, Springer-Verlag,  New York Berlin Heidelberg

\bibitem[\protect\citeauthoryear{Aulanier et al.}{2012}]{aulanier} Aulanier, G., Janvier, M., \& Schmieder, B.\ 2012, A\&A, 543, 110

\bibitem[\protect\citeauthoryear{Barry and Hartigan}{1993}]{b20} Barry D., Hartigan J. A., 1993, Journal of the American Statistical Association, 88, 421, 309

\bibitem[\protect\citeauthoryear{Bessell et al.}{1998}]{bessell} Bessell, M.~S., Castelli, F., \& Plez, B.\ 1998, AAP, 333, 231

\bibitem[\protect\citeauthoryear{Brown et al.}{2011}]{brown} Brown T. M. et al., 2011, ApJ, 142, 112
\bibitem[\protect\citeauthoryear{Buchhave et al.}{2010}]{Buchhave:10} Buchhave, L.\ A., Bakos,  G.\ {\'A}., Hartman, J.\ D., et al.\ 2010, ApJ, 720, 1118

\bibitem[\protect\citeauthoryear{Cutri et al.}{2003}]{cutri} Cutri, R.~M., Skrutskie, 
M.~F., van Dyk, S., et al.\ 2003, VizieR Online Data Catalog, 2246, 0

\bibitem[\protect\citeauthoryear{Daley,Vere-Jones}{2003}]{b10} Daley D.J,  Vere-Jones D., An Introduction to the Theory of Point Processes, 2003, Springer-Verlag,  New York Berlin Heidelberg

\bibitem[\protect\citeauthoryear{F\H{u}r\'esz}{2008}]{Furesz:08} F\H{u}r\'esz, G. 2008, PhD thesis, Univ. Szeged, Hungary

\bibitem[\protect\citeauthoryear{Gregory \& Loredo}{1992}]{GL1992} {Gregory}, P.~C. \& {Loredo}, T.~J. 1992, ApJ, 398, 146

\bibitem[\protect\citeauthoryear{Jenkins et al.}{2010}]{jenkins} Jenkins J.~M., Caldwell D.~A., Chandrasekaran H., et al.\ 2010, ApJ, 713, L120 

\bibitem[\protect\citeauthoryear{Kitze et al.}{2014}]{kitze} Kitze M., Hambaryan V.V., Neuh\"auser R., Ginski C., 2014, MNRAS, 442, 3769

\bibitem[\protect\citeauthoryear{Kretzschmar}{2011}]{kretzschmar} Kretzschmar M., 2011, A\&A, 530, 84

\bibitem[\protect\citeauthoryear{Kutoyants}{1980}]{b7} Kutoyants Y.A., Estimation of Parameters of Stochastic Processes (in russian), {1980}, Armenian Academy of Science, Yerevan

\bibitem[\protect\citeauthoryear{Latham et al.}{2002}]{Latham:02} Latham, D.\ W., Stefanik, R.\ P., Torres, G., Davis, R.\ J., Mazeh, T., Carney, B.\ W., Laird, J.\ B., \& Morse, J.\ A. 2002, AJ, 124, 1144

\bibitem[\protect\citeauthoryear{Maehara et al.}{2012}]{b5}Maehara H. et al., 2012 ,Nature, 485, 478

\bibitem[\protect\citeauthoryear{Miyake et al.}{2012}]{miyake} Miyake, F., Nagaya, K., Masuda, K., \& Nakamura, T.\ 2012, Nature, 486, 240
 
\bibitem[\protect\citeauthoryear{Neuh{\"a}user \& Hambaryan}{2014}]{neuhaeuser} Neuh{\"a}user, R., \& Hambaryan, V.~V.\ 2014, AN, 335, 949

\bibitem[\protect\citeauthoryear{Neuh{\"a}user \& Neuh{\"a}user}{2015}]{rne} Neuh{\"a}user, R., \& Neuh{\"a}user, D.~L.\ 2015, Astronomische Nachrichten, 336, 225 

\bibitem[\protect\citeauthoryear{Nordstr\"om et al.}{1994}]{Nordstrom:94} Nordstr\"om, B., Latham, D.\ W., Morse, J.\ A., Milone, A.\ A.\ E., Kurucz, R.\ L., Andersen, J., \& Stefanik, R.\ P. 1994, AAP, 287, 338

\bibitem[\protect\citeauthoryear{Pinsonneault et al.}{2012}]{pinsonne} Pinsonneault, M.~H., An, D., Molenda-{\.Z}akowicz, J., et al.\ 2012, ApJ, 199, 30

\bibitem[\protect\citeauthoryear{Ram{\'{\i}}rez \& Mel{\'e}ndez}{2005}]{Ramirez:05} Ram{\'{\i}}rez, I., \& Mel{\'e}ndez, J.\ 2005, ApJ, 626, 465

\bibitem[\protect\citeauthoryear{Schaefer}{1989}]{b2} Schaefer B.E., 1989, ApJ, 337,927

\bibitem[\protect\citeauthoryear{Schaefer}{1991}]{b3} Schaefer B.E., 1991, ApJ. 366, L39L42 

\bibitem[\protect\citeauthoryear{Schaefer, King \& Deliyannis}{2000}]{b4}Schaefer B.E., King J.R., Deliyannis C.P., ApJ.,529, 1026

\bibitem[\protect\citeauthoryear{Schaefer}{2012}]{b1} Schaefer B.E., 2012, Nature, 485, 456

\bibitem[\protect\citeauthoryear{Schrijver et al.}{2012}]{schrijver} Schrijver, C.~J., Beer, J., Baltensperger, U., et al.\ 2012, Journal of Geophysical Research (Space Physics), 117, A08103 

\bibitem[\protect\citeauthoryear{Shibayama et al.}{2013}]{shiba} Shibayama, T., Maehara, H., Notsu, S., et al.\ 2013, ApJ, 209, 5

\bibitem[\protect\citeauthoryear{Skrutskie et al.}{2006}]{skrutskie} Skrutskie, M.~F., 
Cutri, R.~M., Stiening, R., et al.\ 2006, ApJ, 131, 1163

\bibitem[\protect\citeauthoryear{Sudarsky}{2003}]{sudarsky} Sudarsky, D.\ 2003, 
Scientific Frontiers in Research on Extrasolar Planets, 294, 499 

\bibitem[\protect\citeauthoryear{Usoskin et al.}{2013}]{usoskin} Usoskin, I.~G., Kromer, B., Ludlow, F., et al.\ 2013, A\&A, 552, LL3
 
\bibitem[\protect\citeauthoryear{Wichmann et al.}{2014}]{wich} Wichmann, R., Fuhrmeister, B., Wolter, U., \& Nagel, E.\ 2014, A\&A, 567, AA36 

\end{thebibliography}
\end{document}